\documentstyle[preprint,aps,prl]{revtex}

\begin{document}
\draft
\preprint{yan.tex}
\title{Numerical observation of non-axisymmetric vesicles in fluid membranes}
\title{Numerical observation of non-axisymmetric vesicles in fluid membranes}
\author{Yan Jie$^{1}$, Liu Quanhui$^{1}$, Liu Jixing$^1$, Ou-Yang      Zhong-can$^{1,2}$}
\address{
$^{1}$Institute of Theoretical Physics, Chinese Academy of Sciences,\\ 
P.O. Box 2735, Beijing 100080, China\\
$^{2}$Center for Advanced Study, Tsinghua University, Beijing 100084, China}
\date{\today}

\maketitle

\begin{center}{\bf Abstract}\end{center}

{By means of Surface Evolver (Exp. Math,{\bf 1},141 1992), a software package of brute-force energy minimization over a triangulated surface developed by 
the geometry center of University of Minnesota, we have numerically searched the non-axisymmetric shapes under 
the Helfrich spontaneous curvature (SC) energy model. 
We show for the first time there are abundant mechanically stable non-axisymmetric vesicles in SC model, including regular 
ones with intrinsic geometric symmetry and complex irregular ones. We report in this paper several interesting shapes including 
a corniculate shape with six corns, a 
quadri-concave shape, 
a shape resembling {\sl sickle cells}, and a shape resembling {\sl acanthocytes}. 
As far as we know, these shapes have not been theoretically obtained by any curvature model before. 
In addition, the role of the spontaneous curvature in the formation of irregular crenated vesicles has been studied. The results shows a positive spontaneous 
curvature may be a necessary condition to keep an irregular crenated shape being mechanically stable. 

\pacs{PACS numbers: 87. 22. Bt, 62. 20. Dc, 02. 60. -x}

\narrowtext

\section*{I. INTRODUCTION}
Vesicles are bags of lipid bilayer membranes which form spontaneous in aqueous environment under appropriate conditions. In order to study 
theoretically the morphology of vesicles, a lipid bilayer which has liquid crystalline structures and characteristics have long been 
considered as a model. The first description of fluid membrane by a curvature energy model was given by Canham[1], in which 
the local energy density of the form $~(2H)^2$ was introduced, where $H$ is the mean curvature of the surface.
From the current 
prospective, this energy model is a faithful description of a vesicle which consists of a symmetric bilayer. However, real lipid bilayers are not symmetric 
and thus there is not a genuine physical realization of this model. Helfrich [2] proposed from curvature elastic theory in liquid crystal 
the well-known spontaneous curvature (SC) energy model,
in which the energy functional is 
\begin{equation}  
F={\frac{1}{2}} \kappa_c\int (C_1+C_2-C_0)^2 dA+\Delta P\int dV+\lambda \int dA,
\end{equation}
where $dA$ and $dV$ are the surface area and the volume elements, respectively,$\kappa_c$ is an 
elastic modulus,$C_1$ and $C_2$ are the two principal curvatures, and $C_0$ is the spontaneous 
curvature which describes the possible asymmetry of the bilayer membrane. Nonzero values of the spontaneous curvature results from the fact that a lipid bilayer may have a tendency 
to curve one way or the other, due, for example, either to intrinsic chemical asymmetry between the two leaves and/or to a chemical asymmetry between 
the interior and exterior aqueous environments. The Lagrange multipliers $\Delta P$ and $\lambda$ take account of the constraints 
of constant volume and constant area, which can be physically understood as the osmotic pressure between the ambient and the internal environment, and the 
tensile coefficient, respectively. 
Based on the model, many works have been done in the axisymmetric case. In their pioneering work on the 
model, Deuling and Helfrich [3-4] numerically found a catalog of possible axisymmetric vesicles shapes. In their papers they used the nomenclature 
developed in red blood cells to describe the shapes they found. Among these shapes were prolate and oblate ellipsoid, 
stomatocytes, and discocytes, some of which resemble the shapes of human red blood cells (RBCs). 

Locating different branches of shapes of minimal energy in the parameter space, then the division of the 
parameter space represents the so-called phase diagram. In SC model, Seifert et al [5] calculated the two-dimensional phase diagram for axisymmetric shapes in SC model 
within a limited parameter space. The parameters they used are the reduced volume $v$ and the scaled spontaneous curvature $c_0$, 
which are defined as $v={V \over (4/3)\pi R_0^3}$, $c_0=C_0 R_0$, where $R_0$ is defined 
by $A=4 \pi R_0^2$ and $A$,$V$ are the real volume and area of the surface, respectively.

In addition, by performing the variation of the energy functional the general equilibrium shape equation
was derived [6]
   \begin{equation}
    \Delta p+2\lambda H-2k_{c}[2H(H^2-K)+C_{0}K+ (C_0^2/2)H+{\Delta} H]=0,
  \end{equation}
where the operator ${\Delta}$ is the Laplace-Beltrami 
operator $\Delta=({1 \over \sqrt{g}}{\partial}_{i}(g^{ij}\sqrt{g}{\partial}_{j})$, $g$ is the determinant of the 
metric $g_{ij}$ and $g^{ij}=(g_{ij})^{-1}$, $K=C_1 C_2$ is the Gaussian curvature and $H=(1/2)(c_1+c_2)$ is the mean 
curvature (Here we use a different sign convention for $H$ from the original derivation of this general shape equation in [6]). 

Solving the shape equation under the axisymmetric case (the corresponding shape equation can be transformed from (2) into an ordinary differential 
equation [7]), several analytical solutions have been found. Among these 
solutions are the solution of biconcave shape, the torus 
solution, and the beyond Delaunay surfaces [8]. The first two solution have been supported experimentally [9] and the third is 
believed to be a hunt for experiments. All these studies concentrate on axisymmetric vesicle shapes. No axi-symmetric vesicle 
shapes of spherical topology has been reported in SC model. 

On the other hand, in other curvature energy models, some progress has been achieved in finding non-axisymmetric vesicle shapes of spherical 
topology by means of brute-force energy minimization over a triangulated surface. In area difference elastic (ADE) model, 
non-axisymmetric ellipsoid shapes have been reported [10]. very recently, 
in a modified ADE model including the contribution of the compressibility of the total area and volume, Wintz et al [11] reported a catalog of 
starfish shapes. Characteristic for such shapes is their flatness  and 
their multi-fold symmetry. As far as we know, no other non-axisymmetric vesicle shapes have been reported in literature by any curvature 
model.

However, on the experimental side, abundent non-axisymmetric RBC shapes of spherical topology have been observed. They may take very complex 
shapes and many of them even have no any intrinsic geometric symmetry. 
There are a lot of clear images obtained by scanning 
electron microscope 
of RBCs in the book ``Living Blood Cells and their Ultra-structure" [12] including very complex vesicle shapes such as {\sl echinocyte type} 
cells (Fig.98 in [12]) 
which have a characteristic shape with crenations or spicules (nearly) evenly distributed on the surface, {\sl acanthocyte type} 
cells (Fig. 157, Fig. 159 in [12]) which bear a superficial resemblance 
to {\sl echinocytes} but with much fewer spicules irregularly arranged and bent back at their tips, 
{\sl Knizocytes} (Fig 106, Fig. 107 in [12]) which are tri- and quadri-concave shapes, and {\sl Sickle type} cells (Fig.198 in [12]) which show 
a sickle-like shape, et al. These  complex shapes are not yet understood theoretically in the context of bending energy models. 
Some researchers [13] 
guess that such exotic shapes maybe involve other energy contributions such as higher-order-curvature terms and van der Waals attraction of 
the membrane.  However, the conjecture is not so obvious as it seems. Our question is: is it possible that these complex shapes can be described 
by a simple curvature model, e.g., SC model? It is the purpose of this paper to numerically search non-axisymmetric shapes of spherical topology 
in SC model. 

Because of its success in 
finding non-axisymmetric ellipsoidal shape and starfish vesicle, we will also employed the algorithm of brute-force energy minimization over a 
triangulated surface in this study. Such a method directly minimizes 
the total energy. The resulting shape is a local energy minimum, which depends in principle on the initial shape chosen. 

We will also locate each shapes found by us in the phase 
diagram in order to know about in which region of the parameter space these shapes exist. In addition, in order to describe 
vesicle shapes, we will use the same nomenclature as in [12].

The plan of the paper is: Section II describes the algorithm and the procedure; Section III gives the main results; Section IV is the discussion 
and conclusion. 
         
\section*{II. The model, the software and the procedure}

In order to find the the locally stable non-axisymmetric configurations of RBCs, we evaluate the bending energy numerically with constraint of the constant 
volume or constant area based on the SC model. The reason that we don't require the constant volume and constant 
area simultaneously is to avoid possible contradictions of the two constraints in our procedure described below. Under the constraint of constant volume $V$, the parameter 
$\lambda$ is just be understood the tensile coefficient, and, under the constraint of constant area $A$, the parameter $\Delta P$ is 
understood as the osmotic pressure. Since our main purpose is to search possible non-axisymmetric shapes in SC model instead of to study shape transitions, 
such a choice is applicable. To our experience, by our procedure described below the software we used in this study performances better at the constraint of 
constant volume than at that of constant area, so in most cases we adopt the constraint of constant volume.          

The software we used to searching the surfaces is ``Surface Evolver" package of computer programs [14], which is based on a discretization of the curvature 
energy, area, and volume on a triangulated surface. The energy in the ``Evolver" can be a combination of surface tension, gravitational 
energy, squared mean curvature et al. Constraints can be geometrical constraints on vertex positions or constraints on integrated quantities such 
as body volumes, surface area, et al. The constraints are incorporated into the bending energy and the corresponding Lagrange multipliers will be 
reported. The resulting total energy is minimized by a gradient descent procedure. The resulting shape is a local energy minimum. These characteristics 
of the ``Evolver" make it be an useful tool for for studying non-axisymmetric shapes in SC model. In the ``Evolver" the osmotic pressure is denoted 
by an internal pressure $P$, and it can deal with the following energy functional conveniently
\begin{equation}
F=m_1\int (H-H_0)^2 dA+\lambda \int dA-P\int dV,
\end{equation}
where $m_1$ is called the ``weight" of the bending energy. Under the definition of $H=(1/2) (C_1+C_2)$, The model is identical to SC model by the 
transformations: $m_1=2 \kappa_c$, $P=-\Delta P$, and $H_0=C_0/2$. In addition, no particular units of measurement are used in the ``Evolver". In 
order to relate the program values to the real world, then all values should be within one consistent unit system. 

The software has been employed to deal many geometry 
problems such as constant mean curvature surfaces, equilibrium foam structure at al. for several years [15-17], so we can rely on it safely. 
Just as an exercise, we have tested it for the equilibrium condition of a perfect sphere with a given target volume evolved from a cube in the 
SC model. The equilibrium condition for the energy functional (3) is
\begin{equation}
-P r^3+2 \lambda r^2+2m_1H_0r(-1+H_0 r)=0,
\end{equation}
where $r$ is the radius of the sphere,  which is identical to the results in [6]. Under the parameters $m_1=1$, $H_0=1$, 
$\lambda=2$, and the target volume $V=4.189$, we obtained a stable unit sphere with the area $A=12.5774$, and the lagrange multiplier $P=4.0023$, 
which is obviously satisfied the equilibrium condition.  
 
On the other hand, the resulting shape generated by such an algorithm depends 
in principle on the initial shape one chooses. The dependence on the initial shape leads to a real difficulty to choose appropriate initial shape 
for yielding interesting configurations. In order to find new kinds of 
non-axisymmetric shapes, the strength of the dependence on the initial shape chosen must be reduced by some way.  
We adopt a procedure based on making the target configuration geometrically ``far away" from an initial shape. 
Taking the constant volume constraint as an example,the detailed procedure is illustrated in the following:

i). Under certain values of $k_c$, $C_0$, $\lambda$, get (for example) a stable sphere with a given target value of the volume $V_0$.

ii). Taking the sphere as an initial shape, change the target volume to a value $V_1$ which is far from $V_0$. This is in order 
to break the stability of the sphere and make the target shape not being ``adjacent" to the sphere. It is hoped that such a sudden and big 
change in the volume will trigger a ``random walk" of the surface before it finally stops at a locally stable configuration with the volume 
$V_1$. 

iii). Each new generated shapes and the complex intermediate unstable shapes can also be taken as initial shape, so the procedure can leads to many shapes with no ``correlation" among 
each other. Such a way is capable of finding many different kinds of surfaces from a simple initial shape.

The above procedure have generated many striking shapes, as reported in the next section.

\section*{III. The main results}

By the procedure described in the last section, many striking vesicle shapes were found. Some of shapes qualititavily resemble the RBC shapes 
observed experimentally. In order to describe these exotic shapes, we also adopt the nomenclature devoloped in red blood cells. We want to report 
four types of shape in this paper which are {\sl corniculate} shape, {\sl knizocyte type} shape, {\sl sickle type} shape, 
and {\sl acanthocyte type} shape. We drew our intention to the last kind of shape, because it is strikingly irregular without any geometric 
symmetry and because it shows other interesting features such as ``budding" and ``vesiculation" formation. 
The role of the spontaneous curvature in the formation of {\sl acanthocyte type} shapes is studied, which shows a positive spontaneous curvature 
may be a necessary condition for the formation of such exotic shapes. Each shape is mapped into the 
two dimensional phase diagram (Fig.8) spanned by the reduced volume and the scaled spontaneous 
curvature. 

\begin{center}  {\sl Corniculate} shape \end{center}

Figure 1 shows a corniculate surface with six corns, whose location 
in the phase diagram is ($v=0.93$,$c_0=1.28$) denoted by  ``*1" in Fig.8.  We obtain it from a destabilized sphere induced by osmotic 
pressure, without the constraints of constant area and constant volume. The corresponding paprameters are ($P=-4$, $\lambda=0.5$, 
$H_0=1$, $m_1=0.5$, $A=5.15$, $V=1.02$). This shape apparently has rotational symmetry which is 
identical to the {\sl octahedron} and thus isomorphic to $S_4$, where $S_X$ denote the group of permutations of the set $X$ [18]. Though 
we didn't find the shape 
in [12], we hope it will be a hunt for future experiments. In addition, such a shape maybe indicate the way of formation of 
{\sl echinocyte III} vesicle shapes (Fig.98-Fig.100 in [12]) which have 10-50 corns evenly distributed on a nearly spherical surface. 

\begin{center} {\sl Knizocyte type} shape \end{center}  

{\sl Knizocytes} (Fig 106, Fig. 107 in [12]) are tri- and quadri-concave shapes found in the experiments of RBCs. The shapes denoted by 
Figure 2 is a quadri-concave shapes and bears a resemblance of the experimentally observed shape (Fin.106 in [12]). Its location in the 
phase diagram is ($v=0.84$, $c_0=-1.41$) denoted by  ``*2" in Fig.8. We obtained it by under the constraint of constant volume from 
an initial 
shape of a smaller volume. The corresponding paprameters are ($P=4.09$, $\lambda=2$, 
$H_0=-0.5$, $m_1=1$, $A=25.10$, $V=10$). The shape has rotational symmetry which is identical to the {\sl cube} and also isomorphic to $S_4$ [18]. Such a 
shape may be seen under different circumstances. In fresh blood, it may be observed in certain hemolytic anemias. In addition, if a suspension 
of cells is examined between slide and coverslip and an erythrocyte permitted to adhere to the slide, gentle deformation of the cell by a 
current of liquid in the preparation may produce this appearance [19]. 

\begin{center} {\sl Sickle type} shape  \end{center}

Figure 3 bears a resemblance of the 
{\sl sickle} cells in {\sl echinocytic} forms (Fig.198 in [12]). 
The location in the phase diagram is ($v=0.74$, $c_0=-1.48$) and denoted by ``*3" in Fig.8. 
It is obtained under the constraint of constant volume from an initial shape with much smaller volume. 
The corresponding paprameters are ($P=5.03$, $\lambda=1.75$, 
$H_0=-0.5$, $m_1=1$, $A=27.50$, $V=10$). {\sl Sickle} cell is related to {\sl sickle} 
cell disease, a hereditary abnormality. {\sl Sickle} cells appear when affected blood is exposed to a sufficiently low oxygen tension.The phenomenon 
can also be seen by sealing a preparation between slide and coverslip and waiting a few hours or leaving the blood for 24 to 48 hours in a vessel 
without oxygen [20].  

\begin{center} {\sl Acanthocyte type} shape \end{center}

Figure 4 shows a strikingly complex shape without any intrinsic geometric symmetry. Characteristic for this shape is its irregular 
characteristic shape and the several irregular distributed crenations, which are just the same as the so-called {\sl acanthocyte type} cell 
shapes (Fig. 157, Fig. 159 in [12]) observed experimentally. The 
location of this shape in Fig.8 is ($v=0.56$, $c_0=1.89$) denoted by  ``*4". It is obtained by applying the procedure described in 
section II several times from a simpler initial shape under the constraint of constant volume of 2. The corresponding parameters 
are ($P=8.21$, $\lambda=2.25$, 
$H_0=1$, $m_1=1$, $A=11.24$, $V=2$). The designation {\sl acanthocyte} was given by 
Singer et al.[21] to crenated red cells found in an hereditary  illness now characterized by the absence of beta-lipo-protein and 
serious nervous system alterations. The abnormality appears to develop during the lifespan of the cells within the circulation and to be absent or minimal 
in the youngest cells [22]. 
Another interesting point is that this shape shows  clear  ``budding" and ``vesiculation" formation, where the term ``vesculation" has the same definition 
in [23], which distinguishes the (singular) limit at which the radius of the neck connecting the mother and the daughter vesicles 
becomes microscopic. Such abnormal shapes were found abundant in this study. In addition, the formation of such irregular shapes and ``budding" as 
well as  ``vesiculation" seems closely 
related to the spontaneous curvature. We show this by the following several shapes evolved from figure 4 by changing the value of the spontaneous 
curvature. Figure 5 is the result by increasing the value of $H_0$ to $H_0=8$, from which one can see that all the buds and the 
crenations are evolved into ``vesiculations". The location of this figure in Fig.8 is ($v=0.70$, $c_0=14.1$) which is beyond the region 
of the phase diagram.  
The corresponding paprameters are ($P=182.00$, $\lambda=2.25$, 
$H_0=8$, $m_1=1$, $A=9.76$, $V=2$). Figure 6 is the result evolved from figure 4 by setting $H_0=0$, which is a perfect sphere 
with ($P=5.75$, $\lambda=2.25$, 
$H_0=0$, $m_1=1$, $A=7.68$, $V=2$). The location in Fig.8 is ($v=1.00$, $c_0=0$) denoted 
by ``*6", which is a perfect sphere. One can also check the parameters by the equilibrium condition of a sphere in 
SC model.  Figure 7 is the result evolved from figure 4 by setting $H_0=-2$, which is a pear like shape with 
($P=24.88$, $\lambda=2.25$, 
$H_0=-2$, $m_1=1$, $A=8.31$, $V=2$). The location in 
Fig.8 is ($v=0.89$, $c_0=-3.25$) denoted by  ``*7".

\section*{IV. DISCUSSION AND CONCLUSIONS}

The algorithm used in this study has powerful ability to find complex vesicle shapes, providing that the dependence in the initial shape chosen 
can be broken by some way. Though our procedure used in the study can not completely break the dependence in the initial shape, many strikingly 
complex shapes in SC model have been found. Some shapes we searched bear the resemblance of experimentally observed RBC shapes. 
Apparently the procedure is also useful in searching new shapes of high genus and can be used in other curvature models. Though the shapes provided 
in this paper are all new, We have the most strong 
impression of the existence of irregular {\sl acanthocyte type} in SC model, because this is the first irregular shapes described by a simple bending 
energy model. Our study shows that adding new energy contributions, such as higher-order-curvature terms and van der Waals attraction of 
the membrane, is not necessary to account for such anomous shapes. In fact, taking Figure 4 as the initial shape, we have obtained several other 
irregular shapes which are not included in this paper. 

The reason why these shapes have not been reported by other researchers also employing the same algorithm is simple: by gradually changing the control 
parameters such as the reduced volume, the spontaneous curvature, et al. from a (simple/regular) initial shape, the resulting set of shapes obtained 
are all strongly correlated to each other, i.e., all of them belong to a "bifurcation tree" in the phase diagram. However, there may have 
coexisting stable configurations of different symmetry corresponding to a set of control parameters, each of them also leads to 
its own ``bifurcation tree" in the phase diagram. Thus, a "jump" between these ``trees" is required to find complex shapes.   

We are very interested in the role of the spontaneous curvature in the formation of irregular crenated shapes like {\sl acanthocytes} 
observed in RBC experiments (e.g., Fig.159, Fig.157, Fig.110 in [12]). The crenations of the {\sl acanthocytes} in fact can be considered as 
the indication of budding formation. In literature, the theoretical study of budding transition has focused on the axisymmetric case. Under the axisymmetric 
condition, the physical origin of the budding transition from a sphere in SC model is easy to be understood. As area increases (or, as volume decreases) an excess area (compared with the area of a sphere with the 
same volume) is available. If this excess area becomes comparable to $4 \pi (2/C_0)^2$, it becomes favorable to shed the excess area in the form of  
a bud with radius compared to $(2/C_0$, which costs very little energy [24]. According to the natural explanation of budding 
transition,  a positive spontaneous curvature should be a necessary condition. Though in our case the shapes are not axisymmetric, the 
basic mechanism of shedding area into buds should be the same. In addition, since heavily distorted shapes correspond to large excess area, 
it is naturally to think that the mechanism of shedding area also 
accounts for the formation of such a heavily distorted shape. Because of the above reasons, 
we guess that a positive spontaneous curvature is also a necessary condition for the formation of {\sl acanthocytes}. The results shown by 
Figure 5, figure 6, and figure 7 support this conjecture. In addition, we have tested the conjecture for several other irregular crenated 
shapes which are not included in this paper, all the results also support the conjecture. 
From the locations of these shapes in Fig.8, our shapes are mainly in the part of higher reduced volume. Fig.4 has the lowest value of $v$ among 
our shapes, which indicates that the most likely region of irregular shapes is in lower part of $v$. It is equivalent to say that a large 
excess area $A(1-v^({2 \over 3})$ favours the formation of heavily distorted shapes, which is appreciable.

\section*{ACKNOWLEDGMENTS}

We are indebted to Professor Ken Brakke, Professor Karsten Grosse-Brauckmann and Professor Rob Kusner for their guidance on the 
software and useful suggestions. We thank Dr. Zhou Haijun, Dr. Zhao Wei for fruitful discussions.
This work is partly supported by the National Natural Science Foundation of 
China.

\begin{figure}
\caption{{\sl corniculate} shape obtained from a destabilized sphere induced by osmotic 
pressure without constraints of constant area and volume. 
The corresponding location in the phase diagram is ($v=0.93$, $c_0=1.28$).}
\end{figure}
\begin{figure}
\caption{A quadri-concave shapes obtained by random searching under the constraint of constant 
volume. Its location in the phase diagram is ($v=0.84$, $c_0=-1.41$).}
\end{figure}
\begin{figure}
\caption{A {\sl sickle} like shape obtained by random searching. The corresponding location 
in the phase diagram is ($v=0.74$, $c_0=-1.48$).}
\end{figure}
\begin{figure}
\caption{{\sl acanthocyte} like shape obtained by random searching under the constraints of 
constant volume. The corresponding location in the phase diagram 
is ($v=0.56$, $c_0=1.89$).}
\end{figure}
\begin{figure}
\caption{The shape is obtained from Fig.4 by increasing the spontaneous curvature. 
All the buds in Fig.4 are evolved into ``vesiculations", 
which shows a large positive 
$C_0$ favours the formation of ``vesiculation". The corresponding location in the phase diagram 
is ($v=0.70$, $c_0=14.1$).}
\end{figure}
\begin{figure}
\caption{This is a perfectly sphere also obtained from Fig.4 by setting a zero spontaneous 
curvature, which shows a non-zero spontaneous curvature may be a condition for the formation 
of {\sl acanthocyte type} shape. 
The corresponding location in the phase diagram is ($v=1.00$, $c_0=0$).}
\end{figure}
\begin{figure}
\caption{The pear-like shape is also evolved from Fig.4 under a negative spontaneous 
curvature, which shows a positive spontaneous curvature may be a necessary condition 
for the formation of the {acanthocyte type} shapes. 
The corresponding location in the phase diagram is ($v=0.89$, $c_0=-3.25$).}
\end{figure}
\begin{figure}
\caption{A schematic copy of the phase diagram in SC model given by Seifert et al. in [5]. 
The location of the shapes found in the paper is denoted by $*n$ where $n$ 
corrsponds to Fig.$n$. In addition, Fig.5 ($v=0.70$, $c_0=14.1$) is not denoted in it because the parameters are beyond the region of 
the phase diagram.}
\end{figure}

\end{document}